\documentclass[onepage,12pt,a4paper,pdftex]{article}

\usepackage{titletoc}
\usepackage{graphicx, amsfonts, amsmath, amssymb, amscd,  theorem, exscale,
epic, eepic, epsfig}

\usepackage{hyperref} %

\usepackage{makeidx}
\makeindex

\newcounter{Figure}
\theoremstyle{plain}
\theoremheaderfont{\scshape}

\newtheorem{The}{\bf Theorem}

\theorembodyfont{\upshape}

\theoremheaderfont{\scshape\small} \theorembodyfont{\scshape\small}

\newcommand{\real}{ {\mathbb R} }

\newcommand{\slap}{\mbox{$ \triangle \mkern -13mu / \ $}}
\newcommand{\nlap}{\mbox{$ \nabla \mkern -13mu / \ $}}
\newcommand{\dlap}{\mbox{$ div \mkern -13mu / \ $}}

\newcommand{\Lbar}{\underline{L}}
\newcommand{\be}{\begin{equation}}
\newcommand{\ee}{\end{equation}}
\newcommand{\bea}{\begin{eqnarray}}
\newcommand{\eea}{\end{eqnarray}}
\newcommand{\beas}{\begin{eqnarray*}}
\newcommand{\eeas}{\end{eqnarray*}}

\begin{document}

\begin{center}
{\Large \bf Gravitational Waves and Their Memory  \\ \vspace{3pt} in General Relativity} \\ 
\end{center}

\vspace{0.1cm}

\begin{center}
{\large \bf Lydia Bieri, David Garfinkle, Shing-Tung Yau} \\ 
\end{center}

\vspace{0.5cm}

{\bf Abstract:} 
General relativity explains gravitational radiation from binary black hole or neutron star mergers, from core-collapse supernovae and even from the inflation period in cosmology. These waves exhibit a unique effect called memory or Christodoulou effect, which in a detector like LIGO or LISA shows as a permanent displacement of test masses and in radio telescopes like  NANOGrav as a change in the frequency of pulsars' pulses. It was shown that electromagnetic fields and neutrino radiation enlarge the memory. Recently it has been understood that the two types of memory addressed in the literature as `linear' and `nonlinear' are in fact two different phenomena. The former is due to fields that do not and the latter is due to fields that do reach null infinity. \\ \\ 
{\small {\bf 2010 Mathematics Subject Classifications:} 35A01, 35L51, 35Q76, 83C05, 83C35. }

\section{Introduction} 
\label{intro}

Gravitational waves occur in spacetimes with interesting structures from the points of view of mathematics and physics. These waves are fluctuations of the spacetime curvature. Recent experiments \cite{bicep2*1} claim to have found imprints in the cosmic microwave background of gravitational waves produced during the inflation period in the early Universe, though it has been argued \cite{spergel1} that foreground emission could have produced the experimental signal. Gravitational radiation from non-cosmological sources such as mergers of binary black holes or binary neutron stars as well as core-collapse supernovae are yet to be directly detected. We live on the verge of detection of these waves and thus a new era of astrophysics where information from formerly opaque regions will be laid open to us. Experiments like LIGO on Earth or LISA planned for space or the pulsar-timing arrays using radioastronomy like NANOGrav are expected to detect gravitational waves in the near future. 
These waves exhibit a memory effect, called the Christodoulou effect, which in the LIGO or LISA experiments manifests itself by displacing test masses permanently and in the radio telescopes like NANOGrav by changing the frequency of pulsars' pulses. The following works by Zel'dovich, Polnarev, Braginsky, Grishchuk and Thorne \cite{zelpol}, \cite{braggri}, \cite{bragthorne} treat a `linear' memory due to a change in the second time derivative of the source's quadrupole moment. This effect was believed to be small. However, Christodoulou \cite{chrmemory} established the `nonlinear' memory effect resulting from energy radiated in gravitational radiation. Christodoulou also showed that his new effect is much larger than the formerly known `linear' one. The first two of the present authors showed \cite{lbdg2}, \cite{lbdg3} that, what in the literature has been known as the `linear' memory is due to fields that do not reach null infinity, and what has been referred to as `nonlinear' memory is due to fields that do reach null infinity. 
Christodoulou \cite{chrmemory} established his result for the Einstein vacuum equations. In general relativity (GR) matter or energy fields on the right hand side of the Einstein equations (\ref{Einst1}) generate curvature. Do these fields contribute to the memory effect? And how large would such a contribution be? 
The first and the third of the present authors with Chen proved \cite{1lpst1}, \cite{1lpst2} that electromagnetic fields enlarge the Christodoulou effect. 
The first and the second of the present authors established \cite{lbdg1} a contribution to memory from neutrino radiation in GR, where the neutrinos are described by a null fluid. 
The same authors also investigated memory for more general energy-momentum tensors \cite{lbdg3}, and they found \cite{lbdg2} an analog to gravitational memory within the regime of the Maxwell equations, which establishes this phenomenon for the first time outside of GR.

Radiative spacetimes are solutions of the Einstein equations 
\be \label{Einst1}
R_{\mu \nu} - \frac{1}{2} g_{\mu \nu} R = 8 \pi T_{\mu \nu}  
\ee
with $R_{\mu \nu}$ denoting the Ricci curvature tensor, $R$ the scalar curvature, $g_{\mu \nu}$ the metric tensor and $T_{\mu \nu}$ the energy-momentum tensor of any fields considered. If the right hand side is not trivial, then corresponding matter equations have to be provided. 
Denote by $(M, g)$ our $4$-dimensional spacetime manifold where $g$ is a Lorentzian metric obeying (\ref{Einst1}). 

Often the left hand side of (\ref{Einst1}) is denoted by $G_{\mu \nu}$ which is called the Einstein tensor. The twice contracted Bianchi identities yield 
\be \label{Bianchi2}
D^{\nu} G_{\mu \nu} = 0 . 
\ee
We see from equations (\ref{Einst1}) and (\ref{Bianchi2}) that the energy-momentum tensor $T_{\mu \nu}$ is symmetric and divergence free 
\be \label{T1}
D^{\nu} T_{\mu \nu} = 0 . 
\ee

The Einstein equations (\ref{Einst1}) are of hyperbolic character and exhibit interesting geometric structures. In mathematical GR the methods of geometric analysis have proven to be most effective in solving some of the most challenging burning problems. 
It turns out that a lot of information about radiation can be gained from the study of asymptotics of spacetimes that were constructed within stability proofs. 
Whereas in the latter, the initial data has to fulfill corresponding smallness assumptions, the emerging asymptotic structures are independent of smallness. In fact, these results hold for large data such as binary black hole mergers and other sources of gravitational radiation. We address this connection in sections \ref{Cauchystab} and \ref{rad}.

Many physical cases require to study the Einstein equations in vacuum. 
By setting to zero the right hand side of (\ref{Einst1}) we obtain the 
Einstein vacuum (EV) equations 
\be \label{EV1}
R_{\mu \nu } = 0 \ . 
\ee 
Solutions of the EV equations 
are 
spacetimes $(M, g)$, where $M$ is a four-dimensional, oriented, time-oriented, differentiable manifold and $g$ 
is a Lorentzian metric obeying the EV equations. This is the first of the following matter models. Among the most popular ones we find: 
\begin{enumerate}
\item The Einstein vacuum (EV) case with $T_{\mu \nu} = 0$ and equations (\ref{EV1}). 
\item The Einstein-Maxwell (EM) case with $T_{\mu \nu} = \frac{1}{4 \pi} \big(  F_{\mu}^{\ \rho} F_{\nu \rho} - \frac{1}{4}
g_{\mu \nu} F_{\rho \sigma} F^{\rho \sigma}  \big)$ 
where $F$ denotes the electromagnetic field, thus an antisymmetric covariant 2-tensor. 
$T_{\mu \nu}$ being trace-free, the Einstein-Maxwell system reads 
\bea
R_{\mu \nu} & =  & 8 \pi T_{\mu \nu} \nonumber \\ 
D^{\mu} F_{\mu \nu} & = & 0  \label{EM1*2} \\ 
D^{\mu} \ ^* F_{\mu \nu} & = & 0. \ \   \nonumber 
\eea
\item The Einstein-Yang-Mills (EYM) equations with Lie group $G$ with $T_{\mu \nu}$ being the stress-energy tensor associated to the Yang-Mills curvature $2$-form $F_{\mu \nu}$. 
The latter takes values in the Lie algebra of $G$ and has to satisfy the EYM system. 
$F$ is written in terms of the potential $1$-form $A$. The tensor $T_{\mu \nu}$ is trace-free as in the Maxwell case. But now for non-Abelian groups the YM potential does not disappear from the equations. 
\item Fluid models such as a perfect fluid, in particular a null fluid. In each case $T_{\mu \nu}$ takes a specific form. The obvious conservation laws have to be fulfilled including the particle conservation law. 
\end{enumerate}

Section \ref{hist} gives a brief history of the efforts to detect gravitational waves.
In section \ref{Cauchystab} we introduce the Cauchy problem for the Einstein equations and explain the stability results for asymptotically flat spacetimes. Section \ref{rad} connects these results to gravitational radiation. In part \ref{mem} we investigate the Christodoulou memory effect of gravitational waves for the Einstein equations. Finally, in the last two sections \ref{det} and \ref{cosm} the prospects for the detection of gravitational wave memory is discussed and a brief discussion of gravitational waves in the cosmological setting is given.  

As the Cauchy problem, black holes, stability of the Kerr solution, cosmology and related topics are covered by other articles, we only briefly touch these here and say as much as needed for our purposes. Also, for a complete list of references regarding these topics we refer to the corresponding articles in this volume.

\section{History of Gravitational Wave Observations} 
\label{hist}

Gravitational radiation can be understood in analogy with the more well known case of electromagnetic radiation.  The motion of electric charges makes waves of changing electric and magnetic fields which propagate outward from their sources, travel at the speed of light, carry energy away from their sources, and fall off like $r^{-1}$ at large distances $r$ from the source.  Electromagnetic waves can be detected directly through their effect on the motion of charges, and indirectly by measuring the amount of energy that their sources lose.  Similarly, the motion of masses makes waves of changing spacetime curvature which 
propagate outward from their sources, travel at the speed of light, carry energy away from their sources, and fall off 
like $r^{-1}$ at large distances $r$ from the source.  Gravitational waves can be detected directly through their effect on the motion of objects in free fall, and indirectly by measuring the amount of energy that their sources lose.  The main differences between the two types of radiation are (1) gravitational waves are much weaker, and (2) the equation that describes gravitational radiation is nonlinear.  Because of the weakness of the gravitational interaction, there are no detectable manmade sources of gravitational radiation: any gravity waves that we could produce in the lab would be too small to be detected.  Only astrophysical sources have any hope of producing detectable gravitational waves, and the most promising sources are those in which gravity is strongest and most dynamical: colliding black holes and neutron stars, and supernova explosions.

Despite the crucial nature of nonlinearity in this chapter, we will first consider what happens when we neglect it.  That is we will consider that class of solutions where the gravitational fields are everywhere sufficiently weak that they can be well approximated by neglecting the nonlinear terms in the Einstein field equation.  Then the field equation reduces to a linear equation that is similar to Maxwell's equations for electromagnetic fields.  In addition we further restrict attention to the case where the sources are moving slowly compared to the speed of light.  Then as for Maxwell's equations one obtains an expansion for the radiation field in terms of time derivatives of the multipole moments of the source.\cite{flanagan}  For electromagnetism the conservation of charge forces the first term in this expansion to vanish and so the dominant term involves the dipole moment.  For gravity the conservation of energy and momentum forces the first two terms in the expansion to vanish, and the dominant term involves the quadrupole moment.  In particular, in this weak field slow motion approximation for gravity one obtains the following formula for the rate at which energy is lost in gravitational radiation:
\begin{equation}
P = {\frac 1 5} {{\dddot {\hat Q}}^{ab}}{{\dddot {\hat Q}}_{ab}}
\end{equation} 
Here $P$ is the power radiated in gravitational radiation, ${Q^{ab}} = \int {d^3} x \rho {x^a}{x^b}$ is the mass quadrupole moment tensor, ${\hat Q}^{ab}$ is the trace-free part of $Q^{ab}$ and an overdot denotes derivative with respect to time.    
This is the so called quadrupole formula of general relativity.  As derived within linearized gravity, the quadrupole formula applies only to systems whose motions are caused by non-gravitational forces.  However, there is also a derivation of the quadrupole formula using the post-Newtonian approximation (see e.g. \cite{PoissonWill}) that applies also to systems whose motions are influenced by gravity.
The quadrupole formula immediately suggests a strategy for the indirect detection of gravitational waves: find an astrophysical system whose motion can be measured very accurately.  Measurement of the motion allows one to calculate the quadrupole moment and thus the rate at which general relativity predicts that the system will lose energy.  But that loss of energy will give rise to a change in the motion of the system, which can be measured through prolonged observation of the system.  It is precisely this strategy that was adopted by Hulse and Taylor\cite{taylor} in their discovery and observations of the binary pulsar.  This system consists of two neutron stars in orbit around each other.  One of these neutron stars is a pulsar: a rotating neutron star from which we receive a pulse of radio waves once each rotation.  This series of pulses acts as an extremely accurate clock which also enables measurement of all details of the orbit through Doppler shifts in pulse arrival times from the orbital motion of the neutron star.  As the binary pulsar system loses energy, the size of the orbit shrinks and the period of the orbit gets shorter.  The quadrupole formula for energy loss then yields a prediction for the rate of change of the period.  Hulse and Taylor measured the rate of change of period and found agreement with the prediction of general relativity.  In so doing they have (indirectly) detected gravitational radiation.  

Though the work of \cite{taylor} leaves no doubt about the existence of gravitational radiation, one would still like to directly detect it.  To understand the strategies for direct detection, it is helpful to turn to the properties of geodesics. Objects in free fall follow spacetime geodesics.  The spatial separation between two nearby geodesics will change with time due to the nonuniformity in the gravitational field encoded in the spacetime curvature.  Specifically the separation of geodesics is governed by the geodesic deviation equation
\begin{equation}
{{\ddot S}^i} = - {R_{titj}}{S^j}
\label{geodev}
\end{equation}
Here one geodesic defines the rest frame that determines the time coordinate $t$ and spatial coordinates $x^i$, while 
$S^i$ is the spatial separation between the two geodesics and $R_{\alpha \beta \gamma \delta}$ is the Riemann tensor.
It is instructive to look at the behavior of eqn. (\ref{geodev}) in the weak field slow motion case.  We find
\begin{equation}
{R_{titj}} = {\frac {-1} r} {\cal P} \left [ {{\ddddot Q}_{ij}} \right ]
\label{slomoRiem}
\end{equation}
Here ${\cal P} []$ denotes `projected orthogonal to the radial direction and trace-free.'  While a gravitational wave changes the positions of objects in free-fall, one might have expected that after the wave has passed the objects return to their original positions.  However, using eqn. (\ref{slomoRiem}) in eqn. (\ref{geodev}) and integrating twice with respect to time one obtains
\begin{equation}
\Delta {S^i} = {\frac 1 r} {S^j} {\cal P} \left [ \Delta {{\ddot Q}_{ij}} \right ]
\label{slomomemory}
\end{equation} 
Here $\Delta$ denotes the difference in a quantity between the begining and end of the time of the passage of the gravitational wave.  Thus there is a change in the positions of the free fall objects: a gravitational wave memory.  Note however, that eqn. (\ref{slomomemory}) only holds for slowly moving sources and in the linearized approximation to general relativity.  For this reason, Christodoulou calls this contribution to gravitational wave memory the ``linear memory.''  When one does not make the linear approximation and instead uses the full nonlinear theory of general relativity there is an additional contribution to the memory.  

The geodesic deviation equation immediately suggests two strategies for direct detection of gravitational radiation which we will call ``bar'' and ``beam.''  The geodesic deviation equation implies that the effect of a gravitational wave on a somewhat rigid object (a ``bar'') will be to make the bar vibrate.  This effect will be especially strong if the gravitational wave has a frequency that is close to the resonant frequency of the bar.  One then isolates the bar from all extraneous sources of vibration (to the extent possible) and measures the tiny vibration induced by the passage of gravitational waves.  This is the strategy used by Weber\cite{weber} in his attempt to detect gravitational waves.  Unfortunately, this strategy was not successful, and in hindsight it is easy to see a major weakness of the bar strategy: since the right hand side of eqn. (\ref{geodev}) contains $S^j$, it follows that the signal (change in separation) is proportional to the original separation.  Thus, to get a signal that is sufficiently large to measure, one should start with geodesics that have a sufficiently large separation.  There is a practical limit to the size of the bar and thus a practical limit to the sensitivity of the bar detector.  

This difficulty immediately suggests the alternative strategy: have the detector consist of widely separated objects and keep track of their spatial separation using light (``beams'').  One especially accurate way of using light to measure spatial separation is laser interferometry: a laser interferometer consists of a laser, two mirrors, and a half reflective mirror called a beam splitter, with the mirrors and beam splitter placed in an L shaped configuration.  The laser shines on the beam splitter which splits the beam into two pieces, each of which travels down one arm of the L, reflects off the mirror at the end of the arm and goes back to the beam splitter where the two halves of the beam recombine.  Since the two arms of the L are not exactly equal length, the distances traveled by the two light beams are different.  If this path difference is an odd number of half wavelengths of the laser light then the two light beams will interfere with each other (destructive interference) and the combined beam will be completely dark.  A laser interferometer set at destructive interference allows a very sensitive measurement of any change in the position of the mirrors, since any change in position of a fraction of a wavelength of light will lead to the dark being replaced by a partial beam, and from the brightness of the partial beam one can read off the fraction of the wavelength that the mirrors have moved.  This beam strategy is used by LIGO (Laser Interferometer Gravitational wave Observatory).\cite{riles}  LIGO consists of two detectors (One in Hanford, WA and one in Livingston, LA), each a laser interferometer with a 4 km arm length.  LIGO began operations in 2001 and has not so far detected any gravitational waves.  However, the detector has recently been upgraded (more powerful laser, better mirrors, better shielding of the mirrors from extraneous vibration, etc.).  With the upgrades one can fairly confidently expect the first direct detection of gravitational waves within the next few years.  Other laser interferometers include VIRGO (located in Cascina, Italy with 3 km arm length) and GEO600 (located in Sarstedt, Germany with 600 m arm length).     

If bigger interferometer arm length is better, then one can get it by going to space.  Here the strategy is to use laser interferometry, but with the lasers and mirrors mounted on three satelites in a triangular configuration.  This is the idea behind the proposed LISA (Laser Interferometer Space Antenna) project.  LISA has been variously proposed as a NASA project, an ESA (European Space Agency) project, and as a joint project between NASA and ESA.  It may eventually be built and launched, but not any time in the near future.  Even larger ``arm length'' can be obtained through pulsar timing.  When a gravitational wave passes between a pulsar and the Earth, it distorts spacetime and affects the timing of the arrival of the pulses at the Earth.\cite{detweiler}  If several pulsars are being timed, then the gravity wave will affect the timing of all their pulses, and by an amount that depends on the direction of the line of sight to the pulsar as well as on the direction and the polarization of the gravitational wave.  Thus, the signature of a gravitational wave is a change in the timing of several pulsars with a particular angular pattern.  This is the idea behind the NANOGrav (North American Nanohertz Observatory for Gravitational waves) project.  Radio astronomers have done pulsar timing for many years, but recently there has been a concerted effort to use pulsar timing to detect gravitational waves.  Here the more pulsars that are timed and the longer they are observed, the greater is the overall sensitivity to a gravitational wave signal.

\section{The Cauchy Problem and Stability} 
\label{Cauchystab}

The Cauchy problem for the Einstein equations is discussed first. Then stability results are explored.

\subsection{The Cauchy Problem} 
\label{Cauchy}

The Einstein equations split into a set of constraint equations which the initial data has to satisfy and a set of evolution equations. 
Depending on the matter and energy present on the right hand side of (\ref{Einst1}) 
we also specify corresponding equations for these fields. The goal is to construct a spacetime by solving an evolution problem of the Einstein equations. 
An {\itshape initial data set} consists of a $3$-dimensional manifold $H$, a complete Riemannian metric $\bar{g}$, a symmetric $2$-tensor $k$ and a well specified set of initial conditions corresponding to the matter-fields. 
These have to satisfy the constraint equations. 

A {\itshape Cauchy development} of an initial data set is a globally hyperbolic spacetime $(M, g)$ satisfying the Einstein equations together with  
an imbedding 
$i : H \to M$ 
such that $i_* (\bar{g})$ and $i_* (k)$ are the first and second fundamental forms of $i(H)$ in $M$.

Most commonly used are two settings of foliations. In the first setting, the spacetime is foliated by a 
maximal time function $t$ and an optical function $u$, respectively. The time function $t$ 
foliates our 4-dimensional spacetime into 3-dimensional spacelike hypersurfaces $H_t$, being 
complete Riemannian manifolds. Whereas the optical function $u$ induces a foliation 
of $(M,g)$ into null hypersurfaces $C_u$, which we shall refer to as null cones. 
The intersections $H_t \cap C_u = S_{t,u}$ are 2-dimensional compact Riemannian manifolds.  
In the second setting, the spacetime is foliated by an optical function $u$ and by a conjugate optical function $\underline{u}$. 
The resulting outgoing and incoming null hypersurfaces constitute the double-null foliation. The characteristic initial value problem then 
consists of initial data given on an initial null cone instead of a spacelike hypersurface and the corresponding evolution equations. 
In this paper, we write the equations with respect to the first setting. 

The evolution equations in the EV situation read: 
\beas
\frac{\partial \bar{g}_{ij}}{\partial t} \ & = & \ 2 \Phi k_{ij} \\ 
\frac{\partial k_{ij}}{\partial t} \ & = & \  \nabla_i  \nabla_j \Phi 
\ - \ (\bar{R}_{ij} \ + \ k_{ij} \ tr k \  - \ 2 k_{im} k^m_j) \Phi 
\eeas
whereas the constraint equations are given as: 
\beas
\nabla^i  k_{ij} \ - \ \nabla_j \ tr k \ &  = & \ 0  \\
\bar{R} \ + \ (tr k)^2 \ - \ |k|^2 \ & = & \ 0 \ . 
\eeas
Here, $\Phi$ denotes the lapse function $\Phi := (- g^{ij} \partial_i t \partial_j t )^{- \frac{1}{2}}$, and ${\nabla}$ is the covariant derivative on $H$. Moreover, barred curvature quantities denote the corresponding curvature tensors in $H$. 
If the time function $t$ is maximal, then $tr k = 0$. 
The time vector field $T$, thus the 
future-directed normal to the foliation is given by $T^i =  - \Phi^2 g^{ij} \partial_j t$ and $Tt = T^i \partial_i t = 1$. 
The components of the inverse metric are $g^{ij} = (g^{-1})^{ij}$. 
The metric reads $g = - \Phi^2  dt^2 + \bar{g}$. 
The foliations by $t$ and $u$ supply us with natural vectorfields to work with. 
Consider $S_{t,u}$ that we introduce above as the intersection of $H_t$ and $C_u$.
Let $N$ be the spacelike unit normal vector of $S_{t,u}$ in $H_t$. Let $\{ e_a \}_{a=1,2}$  be
an orthonormal frame on $S_{t,u}$. 
Then 
$(T, N,e_2, e_1)$ is an orthogonal frame. 
Also, let $\hat{T} = \Phi^{-1}T$ denote the unit future-directed time vector, then consider the pair of null normal vectors to $S_{t,u}$, namely $L= \hat{T} +N$ and
$\underline{L}=  \hat{T} -N$. Together with $\{ e_a \}_{a=1,2}$, they 
form a null frame.

In this article, we consider asymptotically flat spacetimes in GR. 
These describe isolated gravitating systems such as galaxies, neutron stars or black holes. 
These are solutions where $(M,g)$ approaches flat Minkowski space with 
diagonal metric $\eta = (-1, +1, +1, +1)$ in the limit of large distance from a spatially compact region. 
Thus, they can be thought of as having 
an asymptotically flat region outside the 
support of the matter.

Consider an 
{\itshape asymptotically flat initial data set}, i.e. 
outside a sufficiently large compact set $K$, $H \backslash K$ is diffeomorphic to the complement of a closed ball in $\real^3$ 
and admits a system of coordinates in which 
$\bar{g} \to \delta_{ij}$ and $k \to 0$ fast enough.

On the way to the well-posedness of the Cauchy problem for the Einstein equations, we first cite de Donder's \cite{dedonder} work on harmonic gauge. 
Several authors then contributed in various ways such as Friedrichs, Schauder, Sobolev, Petrovsky, Leray and others, which led to 
the existence and uniqueness theorems for general quasilinear wave equations. In \cite{leray} Leray formulated global hyperbolicity. 
Finally, the first proper formulation of the Cauchy problem for the Einstein vacuum equations is due to Y. Choquet-Bruhat. 
In 1952, Y. Choquet-Bruhat \cite{bru} treated the Cauchy problem for the Einstein equations locally in time establishing existence and uniqueness of solutions, reducing the Einstein equations to wave equations, introducing wave coordinates. 
The local result led to a global theorem proved by Y. Choquet-Bruhat and R. Geroch in \cite{bruger}, 
stating the existence of a unique maximal future development for each given 
initial data set. 
As a logical further question arises: Is this maximal future development complete? No, not necessarily. This answer by R. Penrose was formulated in his incompleteness theorem. The latter says that `generic' initial data containing a closed trapped surface will yield a spacetime $M$ which is not null geodesically complete (towards the future), 
in particular it will have a maximal future development which is incomplete. 
A {\itshape closed trapped surface $S$} in a non-compact Cauchy hypersurface $H$ is a 
two-dimensional surface in $H$, bounding a compact domain such that 
\[
tr \chi \ < \ 0 \ \ \ \mbox{on } S  \ . 
\]
$\chi$ is the second fundamental form of a spacelike surface $S$ with respect to the outgoing null hypersurface $C$ properly defined in section \ref{rad}. 
The theorem of Penrose and its extensions by S. Hawking and R. Penrose 
led to ask the following: 
Is there any non-trivial asymptotically flat initial data whose maximal development is complete? Yes. 
This answer was given by D. Christodoulou and S. Klainerman \cite{sta} and is discussed in the next section. 
Instead here, we briefly turn to the opposite direction of black hole formation. 

In 2008 D. Christodoulou \cite{chrdmay2008} proved that by the focussing of gravitational waves a closed trapped surface and therefore a black hole will form for initial data that does not contain any closed trapped surface. Since then several authors including S. Klainerman, J. Luk, I. Rodnianski, P. Yu, contributed to this line of research. See the corresponding articles on this topic for detailed information and references, the most recent being \cite{klluro1}. 

Recent regularity results by S. Klainerman, I. Rodnianski and J. Szeftel [\cite{kleinrod1}, \cite{kleinrod2}, \cite{kleinrod3}, \cite{sze1}, \cite{sze2}, \cite{sze3}, \cite{sze4}] proved the bounded $L^2$ curvature conjecture. In particular, the authors showed for solutions of the EV equations that the time of existence depends only on the $L^2$-norm of the curvature and a lower bound of the volume radius of the corresponding initial data set.

\subsection{Stability} 
\label{stab}

In proving the global nonlinear stability of Minkowski space \cite{sta} D. Christodoulou and S. Klainerman achieved the global result. 
Later, a proof under stronger conditions 
for the global stability of Minkowski space for the EV equations 
and asymptotically flat Schwarzschild initial data 
was given by H. Lindblad and I. Rodnianski \cite{lindrod1}, \cite{lindrod2}, the latter 
for EV (scalar field) equations. 
They 
worked with a wave coordinate gauge, showing the wave coordinates to be stable globally. 
Concerning the asymptotic behavior, the results are less precise than the ones 
of Christodoulou-Klainerman [CK] in \cite{sta}. Moreover, there are more conditions to be imposed on the 
data than in \cite{sta}. 
There is a variant for the exterior part of the 
proof from \cite{sta} using 
a double-null foliation by S. Klainerman and F. Nicol\`o in \cite{klnico1}. 
Also a semiglobal result was given by H. Friedrich \cite{fried1} 
with initial data on a spacelike hyperboloid.
N. Zipser \cite{zip}, \cite{zip2} generalizes the CK proof \cite{sta} to the Einstein-Maxwell case. 
L. Bieri \cite{lydia1}, \cite{lydia2} generalizes the CK results \cite{sta} for the Einstein vacuum equations by assuming only $3$ derivatives of the metric to be controlled as opposed to $4$ in \cite{sta} and by assuming that the initial data as $r \to \infty$ behaves like 
\bea
\bar{g}_{ij} = \delta_{ij} + o_3(r^{- \frac{1}{2}}) \nonumber \\ 
k_{ij} = o_3(r^{- \frac{3}{2}}) \label{lbinit1}
\eea 
which relaxes the decay of \cite{sta} by one power of $r$. The latter establishes the borderline decay of the data. 
We recall that Christodoulou-Klainerman in \cite{sta} considered strongly asymptotically flat initial data which as $r \to \infty$ take the form 
\bea
\bar{g}_{ij}  & = &  (1 +  \frac{2M}{r}) \ \delta_{ij}  +  o_4  (r^{- \frac{3}{2}}) \nonumber \\ 
k_{ij}  & = &  o_3  (r^{- \frac{5}{2}})  ,  \label{safk}
\eea
where $M$ denotes the mass. 

Further results in various directions were obtained in \cite{choqubruhcrhusjoiz}, 
\cite{jloi1}, \cite{jloi2}, \cite{jloi3}, and in the cosmological context \cite{HRin}, \cite{csvedberg1}, \cite{jsp1}, \cite{rodsp}, \cite{lukro}. 
For more details the reader may consult the papers on the corresponding topics in the present volume. In this article we concentrate on the stability results and methods which are important for radiation. 
\begin{The} \label{TheCK1} [CK in \cite{sta}] Strongly asymptotically flat initial data $(H_0, \bar{g}_0, k_0)$ of the type (\ref{safk}) which is sufficiently small, yields a unique, causally geodesically complete, globally hyperbolic solution $(M, g)$ of the EV equations. The spacetime $(M, g)$ is globally asymptotically flat. 
\end{The}
The proof of this celebrated result as well as the proofs of \cite{zip}, \cite{zip2} and \cite{lydia1}, \cite{lydia2} use geometric-analytic methods, thus are invariant, in particular they do not depend on any coordinates. All these proofs use the spacetime-foliation introduced above in section \ref{Cauchy} as the first setting, where the foliations are induced by a maximal time function $t$ and an optical function $u$. 
These proofs rely on `the right energies' which we find in the Bel-Robinson tensor. Very roughly the proofs can be summarized in the following steps: 
1) Introduce and estimate the energies, that can be controlled. 
2) Estimate the curvature by a comparison argument with these energies. 
3) Estimate all the geometric quantities within a bootstrap argument by curvature assumptions and evolution equations. Make use of Hodge systems on $S_{t,u}$ coupled to transport equations along the null and spacelike hypersurfaces. 

Note that in the situations of \cite{sta}, \cite{zip}, \cite{zip2} one sees radiation, whereas in the situation of \cite{lydia1}, \cite{lydia2} radiation cannot be seen due to the lower decay of the data. 

The null asymptotic picture of these results is independent from the smallness assumption in the theorems. Whereas smallness guarantees existence and uniqueness of solutions, the asymptotic results hold true for large data as well. Thus, the asymptotics for typical sources of gravitational waves such as merging black holes generating large data can be studied using these results.  

In the next section, we investigate the null asymptotics of radiative spacetimes.

\section{Radiative Spacetimes and Null Infinity} 
\label{rad}

In order for a spacetime to carry radiation, it cannot be spherically symmetric. 
In what follows, we consider spacetimes with asymptotic structures as worked out by Christodoulou and Klainerman in \cite{sta}, by Christodoulou in \cite{chrmemory} for the EV case, as by Zipser in \cite{zip}, \cite{zip2}, by the first and third present authors with Chen in \cite{1lpst1}, \cite{1lpst2} in the EM situation and as by the first and second present authors in \cite{lbdg1}, and by the first author in \cite{lydia5} in the neutrino case. 

The most important quantities to describe radiation are the shears of incoming respectively outgoing null hypersurfaces. These shears $\hat{\chi}$ and $\hat{\underline{\chi}}$ are the 
traceless parts of the corresponding second fundamental forms. The second fundamental form $\chi$ of a spacelike surface $S$ relative to the outgoing null hypersurface $C$ is given by 
$\chi(X,Y) = g(D_X L, Y)$ for any $X, Y$ in $T_pS$ and $L$ generating vectorfield of $C$. Correspondingly, $\underline{\chi}$ is defined in the analogous way with $L$ being replaced by 
$\underline{L}$ the generating vectorfield of the incoming null hypersurface $\underline{C}$.

These shears have limits along each null hypersurface $C_u$ as $t \to \infty$. 
In all the works cited in the first paragraph of this section it is proven for the corresponding situations that 
\be
\lim_{C_u, t \to \infty} r^2 \hat{\chi} = \Sigma (u, \cdot) \ , \ \ \ \ 
\lim_{C_u, t \to \infty} r \hat{\underline{\chi}} = \Xi  (u, \cdot) \ . 
\ee

Versions of the following theorem in radiative spacetimes are proven in the works \cite{sta}, \cite{chrmemory}, \cite{zip}, \cite{zip2}, \cite{1lpst1}, \cite{1lpst2}, \cite{lbdg1}, \cite{lydia5}.  
\begin{The} 
The limit $ \Xi  (u, \cdot)$ behaves like 
\[
\left| \Xi \left( u,\cdot \right) \right| _{\overset{\circ }{\gamma }}\leq
C\left( 1+\left| u\right| \right) ^{-3/2} 
\]
where $| \cdot |_{\overset{\circ }{\gamma }}$ denotes the pointwise norm on $S^2$ with respect to the metric $\overset{\circ }{\gamma }$ being the limit as $t \to \infty$ for each $u$ of the induced metrics on $S_{t,u}$ rescaled by $\frac{1}{r^2}$. 
\end{The}
The behavior of $\Sigma (u, \cdot)$ is given by Theorems 2, 3 and 4. In particular, $\Sigma$ tends to limits $\Sigma^+$ and $\Sigma^-$ with $u \to + \infty$ and $u \to - \infty$ respectively, which is guaranteed by Theorems 2 and 3. Moreover, it follows from Theorem 4 that these limits and in particular the difference $(\Sigma^+ - \Sigma^-)$ are nonzero.

With respect to the $(t, u)$ foliation introduced above we define the following curvature components, with index notation $A, B \in \{ 1, 2 \}$, 
\bea
\alpha_{AB} = R(e_A, L, e_B, L)  &  , & 
\underline{\alpha}_{AB} = R(e_A, \Lbar, e_B, \Lbar)  \nonumber \\ 
\beta_A = \frac{1}{2} R(e_A, L, \Lbar, L) & , & 
\underline{\beta}_A = \frac{1}{2} R(e_A, \Lbar, \Lbar, L)   \label{curvcomp1} \\ 
\rho = \frac{1}{4} R (\Lbar, L, \Lbar, L) & , & 
\sigma = \frac{1}{4}  \ ^* R(\Lbar, L, \Lbar, L) \nonumber 
\eea
In the radiative spacetimes of interest all these components except for $\alpha_{AB}$ and $\beta_A$ converge to their well-defined limits along $C_u$ as $t \to \infty$ for each $u$. The exceptions $\alpha_{AB}$ and $\beta_A$ tend to zero. The most important limit in view of the following discussion is 
\be \label{alphaWlimit1}
\mathcal{A}_W  (u, \cdot) = \lim_{C_u, t \to \infty} r \underline{\alpha} 
\ee
where the index $W$ denotes the Weyl curvature component to be distinguished from curvature generated through matter fields in the Einstein equations. 
When working with the EV equations (\ref{EV1}) the Weyl curvature is equal to the Riemann curvature. However, as soon as fields are present on the right hand side of the full Einstein equations (\ref{Einst1}) they manifest themselves in the energy-momentum tensor $T_{\mu \nu}$ generating extra curvature. Recall that 
the Weyl tensor (which is traceless) is computed from the Riemann-, Ricci- and scalar curvature tensors as 
\[
W_{\alpha \beta \gamma \delta } = 
R_{\alpha \beta \gamma \delta }  - \frac{1}{2}(g_{\alpha \gamma }R_{\beta \delta }+
g_{\beta \delta }R_{\alpha \gamma }-g_{\beta
\gamma }R_{\alpha \delta }-g_{\alpha \delta }R_{\beta \gamma })  + \frac{1}{6}\left( g_{\alpha \gamma }g_{\beta \delta }- 
g_{\alpha \delta }g_{\beta \gamma }\right) R  \ . 
\]
Denote by $T_{C_u}$ the $T_{\Lbar \Lbar}$ component of the energy-momentum tensor 
and its null limit by $\mathcal{T}_{C_u} = \lim_{C_u, t \to \infty} r^2 T_{C_u}$. 
For the EM equations $T_{C_u}$ is a quadratic of the electromagnetic field component which decays like $r^{-1}$ plus extra decay in $u$. 
In the case of a null fluid $T_{C_u}$ again exhibits the decay behavior of $r^{-2}$ plus decay in $u$. 
In all the corresponding cases investigated in \cite{sta}, \cite{chrmemory}, \cite{zip}, \cite{zip2}, \cite{1lpst1}, \cite{1lpst2}, \cite{lbdg1}, \cite{lydia5} there is an interesting relation between the limiting shears and the Weyl curvature limit: 
\begin{The} \label{theArels1}
The curvature component $\mathcal{A}_W$ and the shears $\Xi$, $\Sigma$ are related through 
\[
\frac{\partial \Xi}{\partial u} = - \frac{1}{4} \mathcal{A}_W \ \ , \ \ 
\frac{\partial \Sigma}{\partial u} = - \Xi . 
\] 
Moreover, it is $| \mathcal{A}_W (u, \cdot ) | \leq C (1 + |u|)^{- \frac{5}{2}}$. 
\end{The}

By 
$\Psi$ denote a Ricci rotation coefficient of the null frame, including $tr \chi$, $tr \underline{\chi}$, the shears $\hat{\chi}$, $\hat{\underline{\chi}}$ and the torsion $1$-form 
$\zeta_A  =  \frac{1}{2} g(D_{\Lbar} L, e_A)$ plus its conjugate. By $\Phi$ denote a component of the Weyl curvature and by $T$ a component of the energy-momentum tensor. $\hat{\eta}_{AB}$ is the traceless tensor component of the second fundamental form $k$ tangential to the $S_{t,u}$-surface. $\nlap$ is the covariant derivative on $S_{t,u}$. 
Then it is 
\beas
\nlap_N \hat{\chi} & = & `` \Psi \Psi" + `` \Psi  \hat{\otimes} \Psi" + ``\nlap \hat{\otimes} \Psi" + ``\Phi" + ``T" \\ \\ 
\nlap_N \hat{\eta} & = & `` \Psi \Psi" + `` \Psi  \hat{\otimes} \Psi" + ``\nlap \hat{\otimes} \Psi" + ``\Phi" + ``T" 
\eeas
Propagation equations for $tr \underline{\chi}$ and $tr \chi$: 
\begin{eqnarray}
\frac{dtr\underline{\chi }}{ds}  &=& - \frac{1}{2}tr\chi tr\underline{\chi }  - 2
\underline{\mu }+2\left \vert \zeta \right \vert ^{2} \label{nullstruct1} \\
\frac{dtr\chi }{ds} &=& - \frac{1}{2}\left( tr\chi \right) ^{2}   -\left \vert 
\widehat{\chi }\right \vert ^{2} -  
8 \pi T .   \label{nullstruct2}
\end{eqnarray}
The Gauss equation reads 
\begin{equation*}
K=-\frac{1}{4}tr\chi tr\underline{\chi }+\frac{1}{2}\widehat{\chi }\cdot 
\underline{\widehat{\chi }}- ``W" + \mbox{ contributions from $T$ } 
\end{equation*}%
The null Codazzi and conjugate null Codazzi equations read 
\beas
\dlap \hat{\chi} \ & = & \  - \hat{\chi} \cdot \zeta + \frac{1}{2} (\nlap tr \chi + \zeta tr \chi) - ``W" \\ 
& & + \mbox{ contributions from $T$ } \\ \\ 
\dlap \hat{\underline{\chi}} \ & = & \ \hat{\underline{\chi}} \cdot \zeta + \frac{1}{2} (\nlap tr \underline{\chi} - \zeta tr \underline{\chi}) + ``W" \\ 
& & + \mbox{ contributions from $T$ } 
\eeas
These equations together with other structure equations, which we do not cite, are used to derive radiation laws. 
To start the investigation, we define the Hawking mass 
\[
m(t,u) = \frac{r}{2} + \frac{r}{32 \pi} \int_{S_{t,u}} tr \chi tr \underline{\chi} . 
\]
By taking $\frac{\partial}{\partial u}$ and $\frac{\partial}{\partial t}$ of $m$ and investigating the null limits, we derive the Bondi mass and the Bondi mass loss relation at null infinity. First, we find that the the Hawking mass $m (t,u)$ tends to a limit $M(u)$ called the Bondi mass along each $C_u$ as $t \to \infty$. 
Second, Bondi mass loss formulas are derived. In the EV case \cite{sta}, \cite{chrmemory} this takes the form 
\[
\frac{\partial}{\partial u} M(u) = \frac{1}{8 \pi} \int_{S^2} | \Xi |^2 d \mu_{\overset{\circ }{\gamma } } . 
\]
Whereas in all the other cases \cite{zip}, \cite{zip2}, \cite{1lpst1}, \cite{1lpst2}, \cite{lbdg1}, \cite{lydia5} this formula exhibits an extra term 
\[
\frac{\partial}{\partial u} M(u) = \frac{1}{8 \pi} \int_{S^2} | \Xi |^2  + C \mathcal{T}_{C_u} d \mu_{\overset{\circ }{\gamma } } . 
\]
The positive constant $C$ and the structure of the positive limit $\mathcal{T}_{C_u}$ depend on the specific situation and are investigated in the works cited. 

Defining the function 
\be \label{en1}
\mathcal{E} = \int_{- \infty}^{\infty} | \Xi |^2  + C \mathcal{T}_{C_u} du 
\ee
$\mathcal{E}/8 \pi$ is the total energy radiated away per unit solid angle in a given direction. 
This seeds the basis for section \ref{mem}.

\subsection{Peeling Behavior}
\label{peel}

To understand radiating properties of a spacetime, we need to know the behavior of the curvature at null infinity. Thus decomposing this tensor according to a null frame 
allows one to describe this behavior. However, what is the asymptotic structure? 

Nowadays, one likes to group the answer to this question into the {\itshape Christodoulou-Klainerman picture} and the {\itshape Newman-Penrose picture}.

The null asymptotic behavior of the Riemann curvature tensor $R_{\alpha \beta \gamma \delta}$ is crucial in the present discussion. 
Its specific decay behavior known as ``peeling estimates" was discussed by Bondi et al. \cite{BBM}, Sachs \cite{sachs} and Penrose \cite{pen}. In their pioneering paper \cite{newpen} Newman and Penrose use a null frame to decompose the important geometric quantities. Even earlier \cite{pirani} Pirani finds peeling properties for curvature components. 
These are important for 
spacetimes which can be conformally compactified (see subsection \ref{NPP}).  

In \cite{sta} Christodoulou-Klainerman derive very precise estimates for the Riemann curvature, they obtain peeling up to one component. See Christodoulou's paper \cite{dcinitialvalue} for further details on these spacetimes. (See subsection \ref{CKP}.)
However, global solutions obtained by Klainerman-Nicol\`o \cite{klnico1} when assuming stronger decay and regularity 
were shown to possess peeling estimates \cite{klnico2}.

The {\itshape Newman-Penrose formalism} refers to the structure equations of a null frame coupled with the Bianchi identities.

\subsubsection{Peeling in the Newman-Penrose Picture}
\label{NPP}

Peeling is addressed in the following pioneering papers \cite{BBM}, \cite{newpen}, \cite{pen}, \cite{pirani}, \cite{sachs}. 
If one imposes a smooth conformal compactification of the spacetime, then a specific hierarchy of decay follows for the curvature components with respect to the frame considered. 
In fact, ``smooth" means at least $C^2$. We can think of this conformal compactification as introducing a boundary future null infinity. 
Based on this one can define rescaled Newman-Penrose curvature scalars being functions on this boundary. These are related to rescaled spin coefficients. 
In fact, often in literature the terminology ``spin-coefficient formalism" is used for ``Newman-Penrose formalism". 

Given such a conformal compactification, the curvature components (\ref{curvcomp1}) show the following decay in $r$ as $t \to \infty$ for fixed $u$: 
\beas 
\underline{\alpha} = O (r^{-1})  & , &   
\underline{\beta} = O (r^{-2}) , \\
\rho, \sigma = O(r^{-3}) & , & \\
\beta = O (r^{-4}) & , & 
\alpha = O (r^{-5}) . 
\eeas

 \subsubsection{Peeling in the Christodoulou-Klainerman Picture}
\label{CKP}

In \cite{sta} Christodoulou-Klainerman derive the following behavior in $r$ for the curvature components (\ref{curvcomp1}) as $t \to \infty$ for fixed $u$: 
\beas 
\underline{\alpha} = O (r^{-1})  & , &   
\underline{\beta} = O (r^{-2}) , \\
\rho, \sigma = O(r^{-3}) & , & \\
\beta = O (r^{-{\frac{7}{2}}}) & , & 
\alpha = O (r^{- {\frac{7}{2}}}) . 
\eeas
We note that  in \cite{sta} the components $\alpha$ and $\beta$ of the curvature fail peeling. In fact, the results of \cite{sta} do not allow for a compact but for a complete future null infinity. This behavior is a direct consequence of the stability theorem \ref{TheCK1}. Now one could ask if $\beta$ were to exhibit the stronger fall-off $O (r^{-4})$, if only one asked for more decay of the initial data in theorem \ref{TheCK1}. Christodoulou \cite{dcinitialvalue} shows that this is not the case for generic initial data. 
He \cite{dcinitialvalue} proves that 
at best what one can achieve is $\beta$ behaving like $r^{-4} \log r$. The reason is that the curvature component $\beta$ is coupled to the radiative amplitude, that is to the shear $\underline{\chi}$, and by investigating their limits at null infinity $\beta$ would fulfill peeling only if a certain expression involving the past limit of this shear vanishes. However, it can be computed that for generic inital data 
this very expression cannot vanish. 

From this discussion we conclude that the question of peeling can only be understood from the point of view of a Cauchy problem for the Einstein equations. What assumptions are required on the initial data so that a reasonable definition of future null infinity follows for the emerging spacetime? This indeed is achieved in \cite{sta} establishing the latter asymptotics and thereby defining complete future null infinity.

\section{`Linear' and `Nonlinear' Memory} 
\label{mem}

Gravitational radiation, namely the fluctuation of curvature, manifests itself in the displacements of geodesics. 
Whenever a binary merger or a core-collapse supernova occurs, a gravitational wave packet is sent out which travels at the speed of light along outgoing null hypersurfaces. These waves have two effects on geodesics: First the so-called instantaneous displacements during the `fly by' of the wave packet, second the permanent displacements after passage of the wave packet called the memory. This {\itshape memory} has a {\itshape `linear'} and a {\itshape `nonlinear'} part, the latter being the {\itshape Christodoulou memory}. 

In the relevant works cited in the previous section, it is proven that 
\[
| \Xi | \to 0 \ \  \mbox{as} \ \ |u| \to \infty 
\]
whereas $\Sigma$ takes limits $\Sigma^+$ as $u \to + \infty$ and $\Sigma^-$ as $u \to - \infty$. 
It is also shown that 
\be \label{Sinst1}
\Sigma (u) \ = \ \Sigma^-  \ + \ \frac{1}{2} \ \int_{- \infty}^{u} \Xi (u') \ du' 
\ee
and 
\be \label{Sperm1}
\Sigma^+ \ - \ \Sigma^- \ = \  \frac{1}{2} \ \int_{- \infty}^{\infty} \Xi (u) \ du . 
\ee
${\Sigma (u) - \Sigma^-}$ is related to 
{\itshape instantaneous displacements} and 
${\Sigma^+ - \Sigma^-}$ 
to 
{\itshape permanent displacements} 
of geodesics. 

In \cite{sta}, \cite{chrmemory} for EV, \cite{1lpst1}, \cite{1lpst2} for EM, \cite{lbdg1}, \cite{lydia5} for Einstein equations with neutrino radiation via a null fluid, it is proven that the permanent displacement related to ${\Sigma^+ - \Sigma^-}$ is governed by the energy $\mathcal{E}$. 
Generally, these results can be formulated as follows. 
\begin{The} \label{perm1}
Denote by $\bar{\mathcal{E}}$ the mean value of the energy $\mathcal{E}$ in (\ref{en1}) on $S^2$. 
Denote by $\stackrel{\circ}{{\mbox{$ ( ) \mkern -10mu / \ $}}}$ an operator on $S^2$. Let 
$h$ be the solution with $\bar{h} = 0$ on $S^2$ of the equation 
\[
\stackrel{\circ}{\slap} h = \mathcal{E} - \bar{\mathcal{E}}   . 
\]
Then 
$\Sigma^+ - \Sigma^-$ is determined by the following equation on $S^2$: 
\be \label{Thm**}
\stackrel{\circ}{\dlap} (\Sigma^+ - \Sigma^-) = \stackrel{\circ}{\nlap} h \ \ . 
\ee
\end{The}
Thus, the {\itshape nonlinear memory of gravitational waves} is governed by the energy function (\ref{en1}). 
We see that the shear $\Xi$ contributes, and is the only contribution in the EV case \cite{chrmemory}, but also there are contributions 
through the energy-momentum component $\mathcal{T}_{C_u}$. The contributions from the EM equations are derived in \cite{1lpst1}, \cite{1lpst2}, for 
Einstein-null-fluid and neutrino radiation in \cite{lbdg1} and for a general energy-momentum tensor in \cite{lbdg3}. 

Let us now inquire how these results impact geodesic behavior and the spacetime geometry. 

We consider three geodesics in spacetime and name them 
$\mathcal{G}_0$, $\mathcal{G}_1$, $\mathcal{G}_2$. 
Let $T$ denote the unit future-directed tangent vectorfield of $\mathcal{G}_0$ and 
$t$ the arc length along $\mathcal{G}_0$. 
Moreover, for each $t$ denote by 
$H_t$ the spacelike, geodesic hyperplane through 
$\mathcal{G}_0(t)$ which is orthogonal to $T$. 
Then we consider the {orthonormal frame field} 
$(T, E_1, E_2, E_3)$ along $\mathcal{G}_0$, 
where $(E_1, E_2, E_3)$ is an orthonormal frame for $H_0$ 
at $\mathcal{G}_0 (0)$, parallely propagated along $\mathcal{G}_0$. 
This yields for every $t$ an orthonormal frame  
$(E_1, E_2, E_3)$ for 
$H_t$ at $\mathcal{G}_0 (t)$.
Next, to a point $p$ 
lying in a neighbourhood of 
$\mathcal{G}_0$ we 
assign the 
cylindrical normal coordinates 
$(t, x^1, x^2, x^3)$, based on $\mathcal{G}_0$, if 
$p \in H_t$ and $p = exp X$ with 
$X = \sum_i x^i E_i \in T_{\mathcal{G}_0(t)} H_t$.  
Under assumptions valid on the Earth, one 
replaces the geodesic equation for $\mathcal{G}_1$ and $\mathcal{G}_2$ by the Jacobi equation 
(geodesic deviation from $\mathcal{G}_0$). 
\begin{equation} \label{Jacobi**1}
\frac{d^2 x^k}{d t^2} \ = \ - \ R_{kTlT} \ x^l 
\end{equation}
with 
\[
R_{kTlT} \ = \ R \ (E_k, T, E_l, T) \ . 
\] 
To see better the structures of the curvature terms, we switch to the null vectorfields again where 
$ L = T - E_3$, $\underline{L} = T + E_3$. 
The leading components of the Weyl curvature are  
\bea
\underline{\alpha}_{AB} (W) \ & = & \ R \ (E_A, \ \underline{L}, \ E_B, \
\underline{L}) \\ 
\underline{\alpha}_{AB} (W) \ & = & \ \frac{\mathcal{A}_{AB}(W)}{r} \ + \ o \ (r^{-2}) \ . 
\eea
Whenever fields are present on the right hand side of the Einstein equations (\ref{Einst1}) and they have the 
`right' decay properties we find the leading order component of the energy-momentum tensor to be 
\be \label{nullTlimit1}
T_{\Lbar \Lbar} \  =  \ \frac{\mathcal{T}_{C_u}}{ r^2} + l.o.t. 
\ee
In order to determine displacements of the geodesics, we have to integrate twice equation (\ref{Jacobi**1}). 
Here, it is crucial that we know the relations between the two shears and the Weyl curvature component $\mathcal{A}_W$. 
By virtue of theorem \ref{theArels1} and the geometric-analytic properties of the relevant spacetimes we derive 
the permanent displacement $\triangle x$ as 
\be \label{perm2}
\triangle \ x \ = \ - \ (\frac{d_0}{r}) \ (\Sigma^+ \ - \
\Sigma^-) \ . 
\ee
The right hand side is understood by theorem \ref{perm1} and equation (\ref{en1}). 
The case of the EV equations was investigated by D. Christodoulou \cite{chrmemory}, for the EM equations by L. Bieri, P. Chen, S.-T. Yau \cite{1lpst1}, \cite{1lpst2}, for the Einstein-null-fluid describing neutrino radiation by L. Bieri, D. Garfinkle \cite{lbdg1} and for a general energy-momentum tensor by the latter authors in \cite{lbdg3}.

\section{Detection of Gravitational Wave Memory} 
\label{det}

What are the prospects for detection of gravitational wave memory?  This question is treated in\cite{favata1}.  Here it is important to remember that there has not yet been {\it any} direct detection of gravitational waves.  Therefore it is best to start by considering whether memory is harder to detect than other aspects of gravitational waves.  Because of the weakness of the gravitational force, only the strongest sources can produce gravitational waves that are likely to be detected by present day detectors.  These sources include colliding black holes, colliding neutron stars, and supernova explosions.  The gravitational waves from these sources are not intrinsically weak: the metric perturbation is of order 1 near the source.  However, gravitational waves go like $r^{-1}$ so their strength at the detector depends on the distance between detector and source.  However black hole collisions (and neutron star collisions and supernovae) are very rare events; so rare that one cannot expect even one such event to occur in our galaxy within a reasonable amount of observing time.  Thus when gravity waves are directly detected, the source will most likely be outside our galaxy.  At such a large distance the $r^{-1}$ falloff of the wave will insure that the amplitude of the wave at the detector is extremely small.  The amplitude of the memory is not weaker than that of other aspects of the gravitational wave signal.  However, there are two effects that make other aspects of the gravitational wave signal easier to detect: matched filtering and the properties of seismic noise.  When black holes collide, they do so because they have been in orbit around each other with the orbit becoming ever smaller due to loss of energy from the radiation of gravitational waves, until finally the orbit becomes so small that the black holes merge.  (The same is true for neutron star collisions).  
Through a combination of approximation methods and numerical simulations, predictions are made for the waveform for the inspiral and merging (for both the black hole and neutron star cases).  
The technique of matched filtering is to compare the detector signal to the predicted waveform, with a match between the two constituting a detection.  This is a much more sensitive detection strategy than simply examining the detector signal, since it makes use of the large amount of information in the predicted waveform.  In constrast, memory is essentially only one piece of information and thus not subject to the effective boost in signal that matched filtering provides.  ``Seismic noise'' refers to the fact that even with the best available technology the detector cannot be completely isolated from extraneous vibration.  This vibration is larger at low frequencies and thus makes it more difficult to detect a low frequency gravitational wave.  For these purposes the ``frequency'' of the memory signal is the inverse of the time that it takes the gravitational wave train to pass.  Therefore memory is lower frequency than the other aspects of the gravitational wave signal and is therefore more subject to interference by seismic noise.  These issues of matched filtering and seismic noise are less severe for the case of supernovae than for black hole or neutron star collisions.  However, even in the supernova case memory is harder to detect than other aspects of the gravitational wave.  Thus, one can expect that LIGO will detect gravitational waves before it detects their memory.  Perhaps the best prospect for detecting gravitational wave memory comes from pulsar timing.\cite{levin}  Here the frequency of the gravitational waves to be measured is much lower than for LIGO and the sources of noise are very different.           

\section{Cosmological Waves}
\label{cosm}

In the treatment of this chapter up to now we have considered memory as a property of gravitational waves in an asymptotically flat spacetime, {\it i.e.} one that approaches Minkowski spacetime at large distances from the sources.  However, we live in an expanding universe, described at the largest scale by the Friedmann-Lemaitre-Roberston-Walker (FLRW) metric.  How much difference does that make for the properties of gravitational waves and their memory?  Certainly it makes a large diffence for the waves detected by BICEP2\cite{bicep2*1}.  These waves owe their existence and properties to the expanding universe: they start out as quantum fluctuations of the gravitational field that are then expanded to enormous size during a time of rapid exponential expansion of the very early universe.  And these waves are (indirectly) detected through their effect on the cosmic microwave background, the light left over from the Big Bang explosion.  But what about more ordinary gravitational waves and their memory? such as those produced by colliding black holes and neutron stars and by supernova explosions.  Is the memory from these sources affected by the fact that they are produced in an expanding universe?  One such effect is a redshift: the stretching of the wavelength of the gravitational wave due to the expansion of the universe.  This effect is treated {\it e.g.} in the analysis of \cite{holz}.  However, there may be additional effects of an expanding universe that change the properties of gravitational wave memory.  The treatment of gravitational wave memory in an expanding universe is a current topic of active research of the present authors.

\section*{Acknowledgments}

The authors acknowledge their NSF support. 
LB is supported by NSF grant DMS-1253149 to The University of Michigan. DG is supported by NSF grant PHY-1205202 to Oakland University. 
STY is supported by NSF grants DMS-0804454, DMS-0937443 and DMS-1306313 to Harvard University.    \\ \\ \\ 

 
%
\vspace{10pt}
{\scshape Lydia Bieri \\ 
Department of Mathematics \\ 
University of Michigan \\ 
Ann Arbor, MI 48109, USA} \\ 
lbieri@umich.edu \\ \\ 
{\scshape David Garfinkle \\ 
Department of Physics \\ 
Oakland University \\ 
Rochester, MI 48309, USA } \\ 
garfinkl@oakland.edu \\ \\ 
{\scshape Shing-Tung Yau \\ 
Department of Mathematics \\ 
Harvard University \\ 
Cambridge, MA 02138, USA } \\ 
yau@math.harvard.edu


\begin{thebibliography}{99} 
\bibitem{bicep2*1} BICEP2 Collaboration; Ade, P. A. R; Aikin, R. W.; Barkats, D.; Benton, S. J.; Bischoff, C. A.; Bock, J. J.; Brevik, J. A.; Buder, I.; Bullock, E.; Dowell, C. D.; Duband, L.; Filippini, J. P.; Fliescher, S.; Golwala, S. R.; Halpern, M.; Hasselfield, M.; Hildebrandt, S. R.; Hilton, G. C.; Hristov, V. V.; Irwin, K. D.; Karkare, K. S.; Kaufman, J. P.; Keating, B. G.; Kernasovskiy, S. A.; Kovac, J. M.; Kuo, C. L.; Leitch, E. M.; Lueker, M.; Mason, P.; Netterfield, C. B.; Nguyen, H. T.; O'Brient, R.; Ogburn, R. W., IV; Orlando, A.; Pryke, C.; Reintsema, C. D.; Richter, S.; Schwarz, R.; Sheehy, C. D.; Staniszewski, Z. K.; Sudiwala, R. V.; Teply, G. P.; Tolan, J. E.; Turner, A. D.; Vieregg, A. G.; Wong, C. L.; Yoon, K. W. 
\begin{itshape} BICEP2 I: Detection Of B-mode Polarization at Degree Angular Scales. \end{itshape} 
(2014). 
arXiv:1403.3985 
\bibitem{lydia1} L. Bieri.  
        \begin{itshape} An Extension of the Stability Theorem of the Minkowski Space
in General Relativity. \end{itshape}
        ETH Zurich, Ph.D. thesis.  \textbf{17178}. 
        Zurich. (2007).  
\bibitem{lydia2} L. Bieri.  
        \begin{itshape} Extensions of the Stability Theorem of the Minkowski Space
in General Relativity. Solutions of the Einstein Vacuum Equations. \end{itshape}
        AMS-IP. Studies in Advanced Mathematics. Cambridge. MA. (2009).   
\bibitem{lydia5} L. Bieri.  
        \begin{itshape} Null Fluid Coupled to Einstein Equations in General Relativity.  \end{itshape} 
        Preprint.  
\bibitem{1lpst1} L. Bieri, P. Chen, S.-T. Yau. 
  \begin{itshape} Null Asymptotics of Solutions of the Einstein-Maxwell Equations in
General Relativity and 
  Gravitational Radiation.    \end{itshape} 
  Advances in Theoretical and Mathematical Physics, 15, 4, (2011). 
\bibitem{1lpst2} L. Bieri, P. Chen, S.-T. Yau. 
  \begin{itshape}  The Electromagnetic Christodoulou Memory Effect and its Application to Neutron Star Binary Mergers.   \end{itshape} 
 Class.Quantum Grav. 29, 21, (2012).        
 \bibitem{lbdg2} L. Bieri, D. Garfinkle. 
   \begin{itshape} An electromagnetic analog of gravitational wave memory.    \end{itshape} 
Classical Quantum Gravity \textbf{30}, 195009 (2013). 
 \bibitem{lbdg1} L. Bieri, D. Garfinkle. 
   \begin{itshape} Neutrino Radiation Showing a Christodoulou Memory Effect in General Relativity.   \end{itshape} 
Annales Henri Poincar\'e. 23. 14. 329. (2014). (DOI 10.1007/s00023-014-0329-1). 
\bibitem{lbdg3} L. Bieri, D. Garfinkle. 
  \begin{itshape}   Perturbative and gauge invariant treatment of gravitational wave memory.  \end{itshape} 
Phys. Rev. D. 89. 084039. (2014). 
\bibitem{BBM} H. Bondi, M. G. J. van der Burg and A. W. K. Metzner.  
\begin{itshape} Gravitational Waves in General Relativity. VII. Waves from
Axi-Symmetric Isolated Systems. \end{itshape} 
Proc. Roy. Soc. A. \textbf{269} (1962).  21-52
\bibitem{braggri} V.B. Braginsky, L.P. Grishchuk.  
Zh.Eksp.Teor.Fiz. {\bf 89}, 744. (1985). 
[Sov.Phys. JETP {\bf 62}, 427. (1986)].  
\bibitem{bragthorne} V.B. Braginsky, K.S. Thorne. 
Nature (London) {\bf 327}, 123. (1987). 
\bibitem{bru}  Y. Choquet-Bruhat. 
        \begin{itshape} Th\'{e}or\`{e}me d'existence pour certain syst\`{e}mes d'equations 
        aux d\'{e}riv\'{e}es partielles nonlin\'{e}aires. \end{itshape}
        Acta Math. \textbf{88}. 
        (1952). 141-225. 
\bibitem{choqubruhcrhusjoiz} Y. Choquet-Bruhat, P.T. Chru\'sciel, J. Loizelet. 
\begin{itshape} Global solutions of the Einstein-Maxwell equations in higher dimensions. \end{itshape} 
CQG \textbf{23}. no. 24. (2006). 7383-7394. 
\bibitem{bruger}  Y. Choquet-Bruhat, R. Geroch. 
        \begin{itshape} Global Aspects of the Cauchy Problem in General Relativity.  \end{itshape}
        Comm.Math.Phys. \textbf{14}. 
        (1969). 329-335. 
\bibitem{dcinitialvalue}  D. Christodoulou. 
        \begin{itshape}  The Global Initial Value Problem in General Relativity.  \end{itshape}
The Ninth Marcel Grossmann Meeting (Rome, 2000). ed. by V. G. Gurzadyan et al., World Scientific. Singapore (2002). 44Ð54.
\bibitem{chrmemory}  D. Christodoulou. 
        \begin{itshape} Nonlinear Nature of Gravitation and Gravitational-Wave
Experiments. \end{itshape}
        Phys.Rev.Letters. \textbf{67}. 
        (1991). no.12. 1486-1489. 
\bibitem{chrdlbmathpgrt}  D. Christodoulou.        
        \begin{itshape} Mathematical problems of general relativity theory I and II.
\end{itshape}
         Volume 1: EMS publishing house ETH Z\"urich. (2008). Volume 2 to appear:
EMS publishing house ETH Z\"urich. 
\bibitem{chrdmay2008}  D. Christodoulou. 
        \begin{itshape} The Formation of Black Holes in General Relativity. \end{itshape}
        EMS publishing house ETH Z\"urich. (2009).           
\bibitem{sta} D. Christodoulou, S. Klainerman.
        \begin{itshape} The global nonlinear stability of the Minkowski space.
\end{itshape}
        Princeton Math.Series \textbf{41}. 
        Princeton University Press. Princeton. NJ. (1993). 
\bibitem{damourbl} T. Damour, L. Blanchet. 
Phys.Rev.D \textbf{46}. 10. (1992). 4304-4319.          
\bibitem{dedonder} T. De Donder. 
{\itshape La Gravifique Einsteinienne.} 
Annales de l'Observatoire Royal de Belgique. Brussels. (1921). 
\bibitem{dessart} L. Dessart, C. Ott, A. Burrows, S. Rosswog, and E. Livne
\begin{itshape} Neutrino Signatures and the Neutrino-Driven wind in Binary Neutron Star Mergers \end{itshape}
ApJ \textbf{690} (2009) 1681-1705 
\bibitem{detweiler} S. Detweiler
ApJ {\textbf 234}, 1100 (1979)
\bibitem{epst} R. Epstein. 
Astrophys. J. \textbf{223}, 1037 (1978). 
\bibitem{favata1} M. Favata. 
Astrophys. J. \textbf{696}. L159-L162. (2009). 
\bibitem{flanagan} E. Flanagan and S. Hughes.
\begin{itshape} The Basics of Gravitational Wave Theory. \end{itshape}
New J. Phys. \textbf{7}, 204 (2005)
\bibitem{spergel1} R. Flauger, J. Colin Hill, D. N. Spergel.  
\begin{itshape} Toward an Understanding of Foreground Emission in the BICEP2 Region.  \end{itshape}
(2014). 
arXiv:1405.7351v2
\bibitem{fried1} H. Friedrich. 
        \begin{itshape} On the Existence of $n$-Geodesically Complete or Future Complete Solutions 
        of Einstein's Field Equations with Smooth Asymptotic Structure. \end{itshape}
        Comm.Math.Phys. \textbf{107}. (1986). 587-609.  
\bibitem{jfrau} J. Frauendiener. 
Classical Quantum Gravity \textbf{9}, 1639 (1992).
\bibitem{levin} R. van Haasteren and Y. Levin.
\begin{itshape} Gravitational-wave memory and pulsar timing arrays. \end{itshape}
Mon. Not. R. Astron. Soc. \textbf{401} 2372 (2010)
\bibitem{holz} S. Nissanke, D. Holz, S. Hughes, N. Dalal, and J. Sievers
Astrophys. J. {\textbf 725} (2010) 496-514
\bibitem{klluro1} S. Klainerman, J. Luk, I. Rodnianski.  
\begin{itshape} A Fully Anisotropic Mechanism for Formation of Trapped Surfaces. \end{itshape}
 Inventiones Mathematicae. {\textbf 195}. (2014)
\bibitem{klnico1} S. Klainerman, F. Nicol\`o. 
        \begin{itshape} The Evolution Problem in General Relativity. \end{itshape}
        Progress in Math.Phys. \textbf{25}. Birkh\"auser. Boston. (2003). 
\bibitem{klnico2} S. Klainerman, F. Nicol\`o. 
        \begin{itshape} Peeling properties of asymptotically flat solutions to the Einstein vacuum equations.  \end{itshape}
        Class. Quantum Gravity. \textbf{20}. no. 14. (2003). 3215-3257.          
 \bibitem{kleinrod1} S. Klainerman, I. Rodnianski. 
	\begin{itshape} On the formation of trapped surfaces.        \end{itshape}
(2009) arXiv:0912.5097. To appear in Acta Mathematica.  
 \bibitem{kleinrod2} S. Klainerman, I. Rodnianski, J. Szeftel. 
	\begin{itshape} The bounded L2 curvature conjecture.       \end{itshape}
 (2012) 	arXiv:1204.1767v1.       
 \bibitem{kleinrod3} S. Klainerman, I. Rodnianski, J. Szeftel. 
	\begin{itshape} Overview of the proof of the bounded L2 curvature conjecture.       \end{itshape}
 (2012) arXiv:1204.1772v1.     
\bibitem{leray} J. Leray. 
{\itshape Hyperbolic differential equations.}        
    The Institute for Advanced Study. Princeton. N.J. (1953).     
\bibitem{ottetal} I. Leonor, L. Cadonati, E. Coccia, S. D'Antonio, A. Di Credico, V. Fafone, R. Frey, W. Fulgione, E. Katsavounidis, C. D. Ott, G. Pagliaroli, K. Scholberg, E. Thrane, and F. Vissani.
\begin{itshape} Searching for Prompt Signatures of Nearby Core-Collapse Supernovae by a Joint Analysis of Neutrino and Gravitational Wave Data \end{itshape}
Class. Quantum Grav. \textbf{27}, 084019 (2010)      
\bibitem{lindrod1} H. Lindblad, I. Rodnianski. 
        \begin{itshape} Global Existence for the Einstein Vacuum Equations in Wave Coordinates. \end{itshape}
        Comm.Math.Phys. \textbf{256}. (2005). 43-110.
\bibitem{lindrod2} H. Lindblad, I. Rodnianski. 
        \begin{itshape} The global stability of Minkowski space-time in harmonic gauge.  \end{itshape}
        To appear. 
\bibitem{jloi1} J. Loizelet.         
\begin{itshape} Solutions globales des equqations d'Einstein-Maxwell avec jauge harmonique et jauge de Lorentz. 
   \end{itshape}      
   C.R. Math.Acad.Sci. Paris \textbf{342} no. 7. (2006). 479-482. 
\bibitem{jloi2} J. Loizelet.         
\begin{itshape} Probl\`emes globaux en relativit\'e g\'en\'erale.  
   \end{itshape}      
   PhD thesis. Universit\'e F. Rabelais. Tours. France. (2008). 
  \bibitem{jloi3} J. Loizelet.         
\begin{itshape} Solutions globales des equations d'Einstein-Maxwell. 
   \end{itshape} 
   Ann.Fac.Soc. Toulouse Math. \textbf{6}. no. 3. (2009). 565-610. 
\bibitem{lukro} C. L\"ubbe, J.A. Valiente Kroon.          
\begin{itshape}   A conformal approach for the analysis of the non-linear stability of radiation cosmologies. \end{itshape} 
Ann. Physics 328. (2013). 1Ð25. 
\bibitem{janka} B. M\"uller, H.-T. Janka, and A. Marek
\begin{itshape} A New Multi-Dimensional General Relativistic Neutrino Hydrodynamics Code of Core-Collapse Supernovae III. Gravitational Wave Signals from Supernova Explosion Models \end{itshape}
arXiv:1210.6984v2   
\bibitem{newpen} E.T. Newman and R. Penrose. 
\begin{itshape} An approach to gravitational radiation by a method of spin coefficients.  \end{itshape} 
Journal of Math.Phys. \textbf{3} (1962). 566-578. 
\bibitem{ott1} C. D. Ott.
\begin{itshape} Probing the Core-Collapse Supernova Mechanism with Gravitational Waves. \end{itshape}
Class. Quantum Grav. \textbf{26}, 204015 (2009) 
\bibitem{pen} R. Penrose.       
         \begin{itshape} Zero Rest-Mass Fields Including Gravitation: Asymptotic Behaviour.  \end{itshape} 
         Proc.Roy.Soc. of London. \textbf{A284} (1965). 159. 
\bibitem{pirani} F. Pirani. 
\begin{itshape} Invariant formulation of gravitational radiation theory.  \end{itshape} 
Phys.Rev. \textbf{105} (1957). 1089-1099. 
\bibitem{PoissonWill} E. Poisson and C. M. Will
\begin{itshape} Gravity \end{itshape}
Cambridge University Press, Cambridge, U.K. (2014)
\bibitem{riles} K. Riles
\begin{itshape} Gravitational Waves: Sources, Detectors and Searches \end{itshape}
Prog. Part. Nucl. Phys. \textbf{68} (2013) 1-54           
\bibitem{HRin2} H. Ringstr\"om.    
{\itshape Future stability of the Einstein-non-linear scalar field system.} 
Invent. Math. 173. (2008). 123-208.
\bibitem{HRin} H. Ringstr\"om.          
         \begin{itshape} On the Topology and Future Stability of the Universe. \end{itshape} 
     Oxford Mathematical Monographs. Oxford Univ. Press. (2013). 
\bibitem{rodsp} I. Rodnianski, J. Speck.  
         \begin{itshape} The nonlinear future stability of the FLRW family of solutions to the irrotational Euler-Einstein system with a positive cosmological constant.  \end{itshape} 
J.Eur.Math.Soc. 15. 6. (2013). 2369-2462.     
\bibitem{sachs} R. K. Sachs.  
         \begin{itshape} Gravitational Waves in General Relativity. VIII. Waves in Asymptotically Flat Space-Time. \end{itshape} 
Proc.Roy.Soc. of London \textbf{A270}. (1962). 103-126. 
\bibitem{ksch} K. Scholberg.          
         \begin{itshape} Supernova Neutrino Detection. \end{itshape}
        Ann.Rev.Nuclear and Particle Science. \textbf{62}. (2012). 81-103. 
 \bibitem{ligo} J. R. Smith
\begin{itshape} The Path to the Enhanced and Advanced LIGO Gravitational-Wave Detectors \end{itshape}
Class. Quantum Grav. \textbf{26}, 114013 (2009) 
\bibitem{jsp1} J. Speck. 
\begin{itshape}
The global stability of the Minkowski spacetime solution to the Einstein-nonlinear electromagnetic system in wave coordinates. 
\end{itshape} (2010).	arXiv:1009.6038v3. 
\bibitem{csvedberg1} C. Svedberg.          
         \begin{itshape}  Future stability of the Einstein-Maxwell-Scalar field system and non-linear wave equations coupled to generalized massive-massless Vlasov equations.    \end{itshape} KTH Stockholm. Ph.D. thesis. (2012). 
\bibitem{sze1} J. Szeftel. 
\begin{itshape}
Parametrix for wave equations on a rough background I: Regularity of the phase at initial
time. 
\end{itshape}
(2012)  arXiv:1204.1768v1. 
\bibitem{sze2} J. Szeftel. 
\begin{itshape} Parametrix for wave equations on a rough background II: Construction of the parametrix
and control at initial time. \end{itshape}
(2012)	arXiv:1204.1769v1
 \bibitem{sze3} J. Szeftel. 
 \begin{itshape} Parametrix for wave equations on a rough background III: Space-time regularity of the
phase. \end{itshape} 
 (2012) arXiv:1204.1770v1
 \bibitem{sze4} J. Szeftel. 
  \begin{itshape} Parametrix for wave equations on a rough background  IV: Control of the error term.
  \end{itshape}
(2012)  	arXiv:1204.1771v1
\bibitem{taylor} J.H. Taylor and J.M. Weisberg
\begin{itshape} A New Test of General Relativity - Gravitational Radiation and the Binary Pulsar PSR 1913+16 \end{itshape}
ApJ \textbf{253} (1982) 908-920 
\bibitem{thorne1992} K. S. Thorne.  
Phys. Rev. D \textbf{45}, 520 (1992).
\bibitem{tolwal1} A. Tolish,  R. M. Wald. 
Phys. Rev. D \textbf{89}, 064008 (2014). 
\bibitem{traut} A. Trautman.           
         \begin{itshape} Radiation and boundary conditions in the theory of gravitation.  \end{itshape} 
Bull.Acad.Pol.Sci.Ser.Math.Astron.Phys. \textbf{6} (1958). 407.
\bibitem{truner} M. Turner. 
Astrophys. J. \textbf{216}, 610 (1977). 
\bibitem{vanderBurg}  M. G. J. van der Burg. 
        \begin{itshape} Gravitational Waves in General Relativity X. Asymptotic
Expansions for the Einstein-Maxwell Field \end{itshape}
         Proc. Roy. Soc. A. \textbf{310} (1969).  221-230  
\bibitem{weber} J. Weber.
Phys. Rev. Lett. {\textbf 24} (1970) 276-279    
\bibitem{wiswill} A. G. Wiseman, C. M. Will. 
Phys. Rev. D \textbf{44}, R2945
(1991).            
\bibitem{zelpol} Ya.B. Zel'dovich, A.G. Polnarev. 
Astron.Zh. {\bf 51}, 30. (1974). 
[Sov.Astron. {\bf 18}, 17. (1974)]. 
\bibitem{zip} N. Zipser. 
        \begin{itshape} The Global Nonlinear Stability of the Trivial Solution of
the Einstein-Maxwell Equations.  \end{itshape}
        Ph.D. thesis. Harvard Univ. Cambridge MA. (2000).          
\bibitem{zip2} N. Zipser.  
        \begin{itshape} Extensions of the Stability Theorem of the Minkowski Space
in General Relativity. - Solutions of the Einstein-Maxwell Equations.
\end{itshape}
        AMS-IP. Studies in Advanced Mathematics. Cambridge. MA. (2009).           
\end{thebibliography}
\end{document}